\documentclass[aps,twocolumn,showpacs,superscriptaddress,prl]{revtex4}
\usepackage{epsfig}

\def\avg#1{\langle#1\rangle}

\def\be{\begin{equation}}       \def\ee{\end{equation}}
\def\bea{\begin{eqnarray}}      \def\eea{\end{eqnarray}}
\def\PRA{Phys. Rev. A~}
\def\PRB{Phys. Rev. B~}
\def\PRL{Phys. Rev. Lett.~}
\def\nn{\nonumber}
\def\pp{\parallel}

\begin{document}

\title{Prediction of quantum stripe ordering in optical lattices}
\author{Congjun Wu}
\affiliation{Kavli Institute for Theoretical Physics, University
of California, Santa  Barbara, CA 93106}
\author{ W. Vincent Liu}
\affiliation{Department of Physics and Astronomy, University of
Pittsburgh, Pittsburgh, PA 15260}
\author{Joel Moore}
\affiliation{Department of Physics, University of California, 
Berkeley, CA 94720}
\affiliation{Materials Sciences Division, Lawrence Berkeley
National Laboratory, Berkeley CA 94720
}
\author{Sankar Das Sarma}
\affiliation{Condensed Matter Theory Center, 
Department of Physics, University of Maryland, College Park, MD 20742} 

\begin{abstract}
We predict the robust existence of a novel quantum orbital stripe order
in the $p$-band Bose-Hubbard model of
two-dimensional triangular optical lattices with cold bosonic atoms.
An orbital angular momentum moment is formed on each site %in the ground state
exhibiting a stripe order both in the superfluid and
Mott-insulating phases.
The stripe order spontaneously breaks time-reversal, lattice translation
and rotation symmetries.
In addition, it induces staggered plaquette bond currents in the superfluid
phase.
Possible signatures of this stripe order in the time of flight
experiment are discussed.
\end{abstract}
\pacs{03.75.Lm, 05.30.Jp, 73.43.Nq, 74.50.+r} 
\maketitle

Cold atomic systems with multiple components,
such as large spin systems, exhibit much richer phase diagrams and
properties than the usual spinless bosons and spin-$\frac{1}{2}$ fermions.
For example, various spinor condensations and spin dynamics
have been investigated \cite{myatt1997,ho1998}.
Similarly, large spin fermions also exhibit 
novel features, including hidden symmetries \cite{wu2003},
quintet Cooper pairing states
\cite{wu2005}, and multiple-particle clustering 
instabilities \cite{wu2005a}.
Motivated by such considerations, but for orbital degeneracy,
we investigate the $p$-band Bose-Hubbard (BH) model
for cold atom optical lattices, finding a novel quantum orbital
stripe phase in triangular $p$-band optical lattices.

In solid state physics, orbital dynamics plays important roles 
in transition metal oxides leading to interesting phenomena, 
such as orbital ordering and colossal magnetoresistence \cite{tokura2000}.
In optical lattices, pioneering experiments
on orbital physics have been recently carried out by Browaeys {\it et. al}
\cite{browaeys2005} by accelerating the lattice of bosons, and 
by Kohl {\it et. al} \cite{kohl2005} by using fermionic Feshbach resonance.
These experiments demonstrate the population of  higher orbital bands,
motivating  our theoretical interest in possible orbital
ordering in optical lattices.
Compared to transition metal oxides where
Jahn-Teller distortions often remove the orbital degeneracy,
optical lattices have the advantage of the lattice rigidity,
and thus the orbital degeneracy is robust.
The $p$-band bosons in the square or cubic lattices,
have received much
attention~\cite{scarola2005,isacsson2005,kuklov2006,liu2006}.
For example, Ref. \cite{isacsson2005} focuses on the
sub-extensive $Z_2$ symmetry \cite{moore2004,xu2004,nussinov2005} and the
resulting nematic superfluid order by considering 
only the $\sigma$-type bonding in the band structure.
By further keeping the $\pi$-bonding term, the ground state
is shown to break time reversal (TR) symmetry spontaneously,
forming an antiferromagnetic order of orbital angular momentum (OAM)
moments \cite{kuklov2006,liu2006}.

The $p$-band bosons in a frustrated optical lattice have, however,
never been studied before.
The experimental realization of the 2D triangular lattice 
has been discussed in the literature \cite{jaksch2005}.
In this paper, we find  a novel quantum 
stripe ordering of the $p$-band bosons in such a lattice.
The onsite Hubbard interaction gives rise to a Hund's rule-like  coupling 
in the OAM channel, resulting  in the formation 
of an Ising OAM moment on each site.
Due to the geometric frustration,
the ground state exhibits a stripe order of the OAM moments which spontaneously
breaks TR, lattice rotation and translation symmetries.
This stripe order bears some superficial similarity to its 
solid-state counterpart observed in  strongly correlated electronic
systems, such as manganites \cite{mori1998},
high T$_c$ cuprates \cite{kivelson1998}, and high-Landau level 
quantum Hall systems \cite{lilly1999},  but is different 
qualitatively since, unlike the solid-state examples, the stripe
order in the $p$-band bosonic triangular optical lattices is fully
quantum in nature and does  not just arise from  the long-range Coulomb
interaction.

We begin with the construction of the $p$-band  BH model 
in a 2D ($x$-$y$) triangular lattice.
The optical potential on each site is approximated by a 3D anisotropic 
harmonic potential with frequencies $\omega_z \gg 
\omega_x=\omega_y=\omega_{xy}$.
Thus we can neglect the $p_z$-band, and 
only consider a two-band model of $p_x$ and $p_y$. 
We define three unit vectors $\hat e_1=\hat e_x, \hat e_{2,3}=
-\frac{1}{2} \hat e_x \pm \frac{\sqrt{3}}{2} \hat e_y$, and 
the two primitive lattice vectors can be taken as
 $a \hat e_{1,2}$ ($a$ the lattice constant).
The projection of the $p_{x,y}$ orbitals along 
the $\hat e_{1,2,3}$ directions are
$ p_1=p_x,  p_{2,3}=-\frac{1}{2} p_x \pm \frac{\sqrt 3}{2} p_y.$
Due to the anisotropic nature of the $p$-orbitals, the hopping
terms are dominated by the ``head to tail'' type $\sigma$-bonding 
(the $\pi$-bonding $t_\perp$ term can be neglected here)
as
\bea
&&H_{0}=t_\pp\sum_{\vec r, i=1,2,3}
\Big \{p^\dagger_{i,\vec r} p_{i,\vec r+ \hat e_i}
+h.c. \Big\},
\label{eq:triHam}
\eea
where $t_\pp$ is positive due to the odd parity of the $p$-orbitals.
The on-site Hubbard repulsion  $H_{int}$  can be calculated from the
contact  interaction of coupling constant $g$:
\bea
H_{int}=\frac{U}{2} \sum_{\vec r}
\Big\{ n^2_{\vec r}-\frac{1}{3} L_{z,\vec r}^2\Big\},
\label{eq:hamint}
\eea
with $n_{\vec r}$  the particle number and 
$L_z=-i(p^\dagger_x p_y -p^\dagger_y p_x)$ the $z$-component OAM;
$U=3g/ [4 (2\pi)^{3/2}\, l_{x} l_y l_z ]$ with $l_{i}=\sqrt{\hbar /
(m\omega_{i})}~ (i=x,y,z)$.
The important feature of $H_{int}$ is its ferro-orbital
nature \cite{liu2006} 
which is analogous to the Hund's rule for atomic electrons.
This implies that bosons on each site prefer to go into the 
the axial states of $p_x\pm i p_y$.
This is because the axial states are spatially more extended
than the polar states $p_{x,y}$, and thus are energetically 
more favorable for $g>0$.

We first consider the weak coupling limit, $U/t\rightarrow 0$.
The Brillouin zone takes the shape of a regular hexagon with the
edge length $4\pi/(3 a)$. 
% {\bf [NOTE: $a_0 \rightarrow a$?! for
% $K_{2,3}$ below are introduced in terms of $a$ and in the TOF figure.]} 
The energy spectrum of $H_0$ is
$E(k)= t_\pp \Big\{ f_{\vec k} \mp \sqrt{f^2_{\vec k}
-3g_{\vec k}} \Big\},$
where $f_{\vec k}= \sum_{i=1}^3 \cos (\vec k \cdot \hat e_i)$
and $g_{\vec k} =\sum_{3\ge i>j\ge 1} \cos(\vec k \cdot \hat e_i) 
\cos (\vec k \cdot \hat e_j)$.
The spectrum contains three degenerate minima  located at
$
K_1=(0, \frac{2\pi}{\sqrt 3 a}), 
K_{2,3}=(\pm\frac{\pi}{a}, \frac{\pi}{\sqrt 3 a})$,
The factor $e^{i \vec K_1 \cdot \vec r}$ takes the value of 
$\pm 1$ uniformly in each horizontal row but alternating
in adjacent rows.
If the above pattern is rotated at angles of $\pm\frac{2\pi}{3}$,
then we arrive at the patterns of $e^{i \vec K_{2,3} \cdot \vec r}$.
Each eigenvector is a 2-component superposition vector of $p_x$ and $p_y$
orbitals.
The eigenvectors at energy minima are
$\psi_{K_1}= e^{i\vec K_1 \cdot \vec r} |p_y\rangle$.
$\psi_{K_{2,3}}$ can be obtained by rotating 
$\psi_1$ at angles of $\pm \frac{2\pi}{3}$ respectively.

\begin{figure}
\centering\epsfig{file=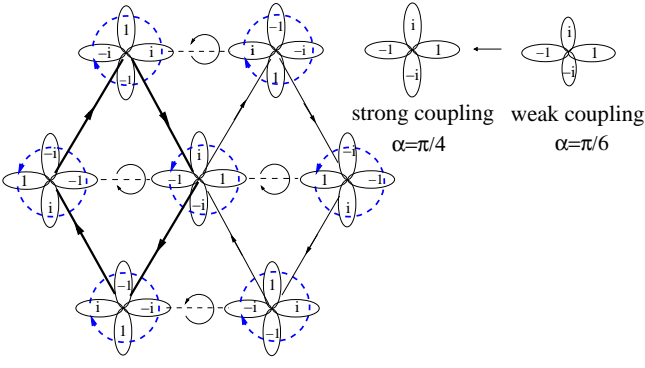,clip=1,width=\linewidth,angle=0}
\caption{The condensate configuration in the real space
described by Eq.(\ref{eq:conden}) with TR, rotation, 
and U(1) symmetry breaking.
The unit cell contains 4 sites as marked with thick lines.
The OAM moments (dashed arrowed circles) form the stripe order, and
induce bond currents exhibiting staggered plaquette moments
(solid arrowed circles).
$\alpha$ in Eq. \ref{eq:conden} depends on the interaction
strength with $\alpha=\frac{\pi}{6} (\frac{\pi}{4})$ 
in the weak (strong) coupling limit respectively.
Currents vanish in the weak coupling limit,
and exist on tilted bonds at finite interaction strength
with directions specified by arrows.
}
\label{fig:triorbcur}
\end{figure}

The ground state  condensate wavefunction $\Psi_c$ can be constructed 
as follows.
Any linear superposition of the three  band minima
$\Psi_c(\vec r)= c_1 \psi_{K_1} +c_2 \psi_{K_2} +c_3 \psi_{K_3}$
with the constraint $|c_1|^2+|c_2|^2+|c_3|^2=1$
equally minimizes the kinetic energy $H_0$.
However, an  infinitesimal $U/t$ removes the band degeneracy 
to further optimize the interaction energy $H_{int}$.
After straightforward algebra,  we find the optimal configurations
occur at
$c_1=0,  c_2=\frac{1}{\sqrt{2}},  c_3=\frac{i}{\sqrt{2}}$,
and
its  symmetrically equivalent partners.
Thus the mean field condensate can be expressed as
$
\frac{1}{\sqrt {N_0 !}}\Big\{\frac{1}{\sqrt 2} (\psi^\dagger_{K_2} +i \psi^\dagger_{K_3})\Big\}^{N_0}
|0\rangle
$
with $|0\rangle$ the  vacuum state and
$N_0$ the particle number in the condensate.
This state breaks the $U(1)$ gauge symmetry, as well as 
TR and lattice rotation symmetries, thus the
ground state manifold is $U(1)\otimes Z_2 \otimes Z_3$.
This state also breaks lattice translation symmetry, 
which is, however, equivalent to suitable combinations of 
$U(1)$ and lattice rotation operations.

For better insight, we transform the above momentum space condensate to 
the real space. The orbital configuration on each site reads
\bea
e^{i\phi_{\vec r}}(\cos \alpha |p_x\rangle +i\sigma_{\vec r}
\sin \alpha |p_y\rangle)
\label{eq:conden}
\eea
with $\alpha=\frac{\pi}{6}$ as $U/t\rightarrow 0$.
The general configuration of $\alpha$  is
depicted in Fig. \ref{fig:triorbcur} for later convenience.
The $U(1)$ phase $\phi_{\vec r}$ is specified at the right lobe
of the $p$-orbital. 
The Ising variable $\sigma_{\vec r}=\pm1$ denotes 
the direction of the OAM, and is represented by the anti-clockwise (
clockwise) arrow on each site.
Each site exhibits a nonzero
OAF moment and breaks TR symmetry.
At $U/t\rightarrow 0$, $p_{x,y}$  are not equally populated, 
and the moment per particle is $\frac{\sqrt{3}}{2} \hbar$.
This does not fully optimize  $H_{int}$
which requires $L_{z,\vec r}=\pm \hbar$.
However, it fully optimizes $H_0$ which 
dominates over $H_{int}$ in the weak coupling limit.
We check that the phase difference is zero along each bond,
and thus no inter-site bond current exists.
Interestingly, as depicted in Fig. \ref{fig:triorbcur},  OAM moments
form a stripe order along each horizontal row.
The driving force for this stripe formation in the SF
regime is the kinetic energy, i.e., the phase coherence between 
bosons in each site.
By contrast, the stripe formation 
in high T$_c$ cuprates 
is driven by the competition between long range repulsion and the
short range attraction in the interaction terms
\cite{kivelson1998}.

Next we discuss the ordering at large large values of $U/t$.
We first minimize $H_{int}$
with $n$ particles per site.
For simplicity, we consider the large $n$ case, then
Hund's rule coupling favors the onsite state
$\frac{1}{\sqrt{n!}}
\Big\{ (\cos \frac{\pi}{4} p^\dagger_x+i \sigma_r \sin \frac{\pi}{4}
p^\dagger_y)\Big\}^n |0\rangle$.
This corresponds to the case $\alpha=\frac{\pi}{4}$ in Fig. 
\ref{fig:triorbcur}.
Because of the anisotropic orientation of the $p$-orbitals, 
the phase difference between two sites along each bond not only 
depends on the $U(1)$ and the Ising variables, but also on 
the direction of the bond as in the $p+ip$ Josephson junction arrays
\cite{moore2004}.
This effect can be captured by a $U(1)$ gauge field.
The effective Hamiltonian then  reads
$
H_\mathrm{eff}=- \frac{1}{2} n t_\pp \sum_{\avg{\vec{r}_1, \vec{r}_2}} 
\cos\big\{\phi_{\vec{r}_1}-\phi_{\vec{r}_2}-A_{\vec{r}_1,\vec{r}_2}(\sigma_{\vec{r}_1},\sigma_{\vec{r}_2}
)\big\} %\nn \\
+ \frac{1}{3} U \sum_{\vec{r}} n^2_{\vec r},
$
where the gauge field $A_{\vec{r}_1,\vec{r}_2}=\sigma_{\vec{r}_1} 
\theta_{\vec{r}_1,\vec{r}_2}-\sigma_{\vec{r}_2}
\theta_{\vec{r}_2,\vec{r}_1}$; $\theta_{\vec{r}_1,\vec{r}_2}$ is the angle
between the bond from $\vec{r}_1$ to $\vec{r}_2$ and the $x$-axis,
and thus $\theta_{\vec{r}_2,\vec{r}_1}=\theta_{\vec{r}_1,\vec{r}_2}+\pi$.
The external gauge flux in the  plaquette $i$
with three vertices $\vec{r}_{1,2,3}$ can be calculated as
\bea
\Phi_i=\frac{1}{2\pi} \sum_{\avg{r,r^\prime}} A_{r,r^\prime}
=\frac{1}{6} (\sigma_{\vec{r}_1}+\sigma_{\vec{r}_2}+\sigma_{\vec{r}_3})
~\mbox{mod 1} .
\label{eq:vort}
\eea
Following the analysis in Ref. \cite{moore2004,castelnovo2004},
the ground state configuration for Ising variables requires 
the flux $\Phi_i$ (vorticity) in each plaquette to be as small as possible,
which are just $\pm \frac{1}{6}$ corresponding to Ising variables
of two $\pm1$'s and one $\mp1$.

\begin{figure}
\hspace{5mm}
\centering\epsfig{file=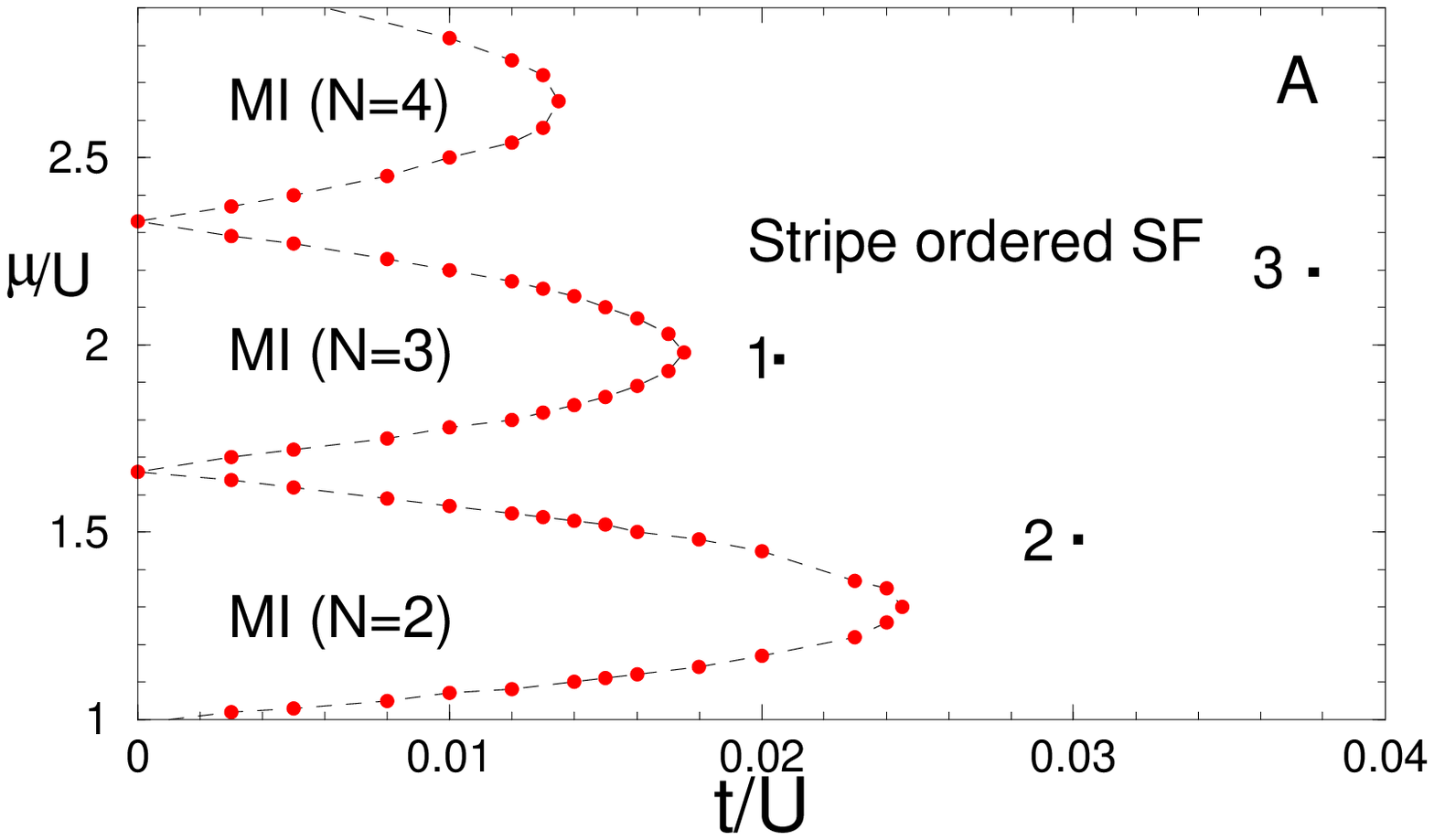,clip=1,width=0.8\linewidth,angle=0}
\vspace{5mm}
\centering\epsfig{file=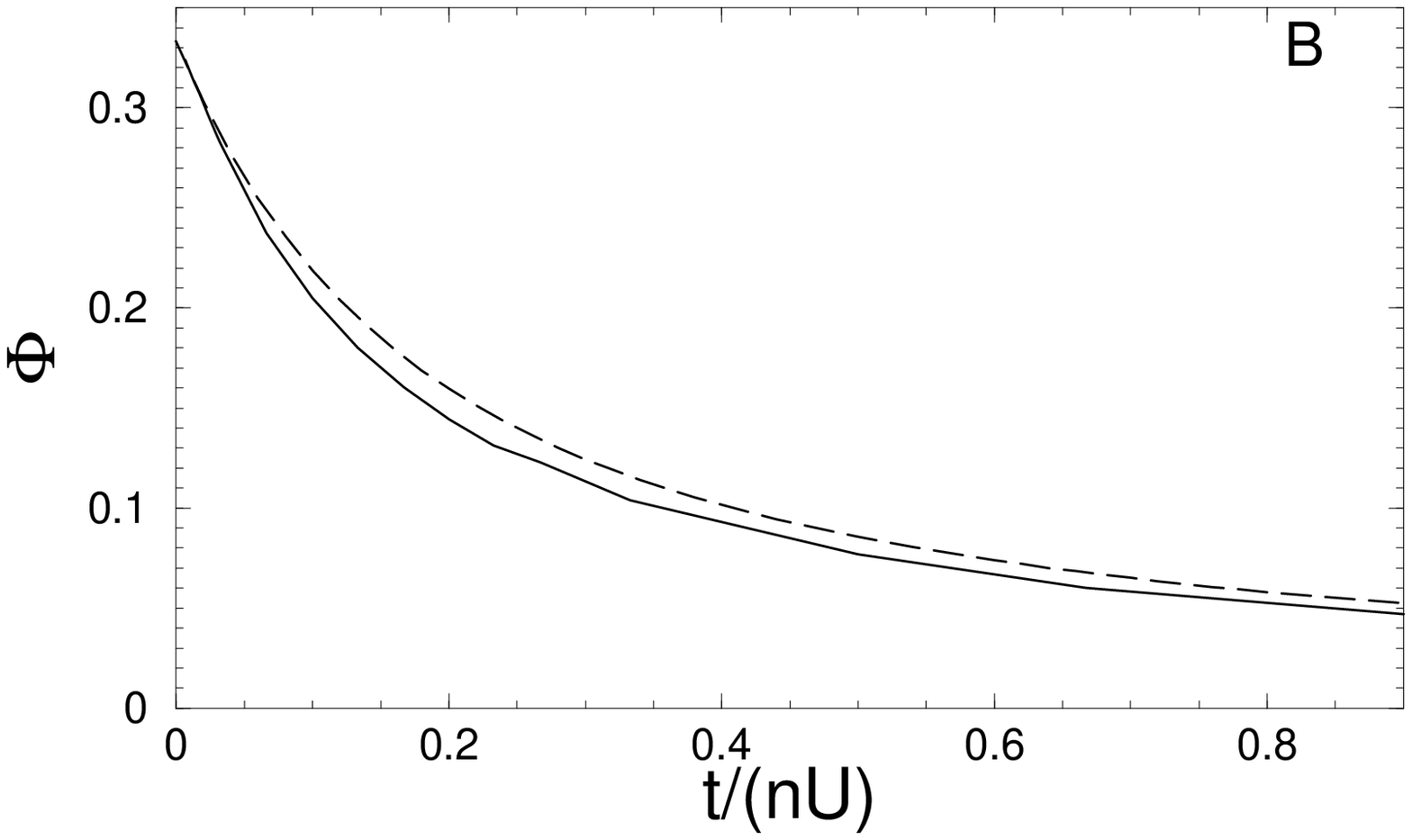,clip=1,width=0.8\linewidth,angle=0}
\caption{A) Phase diagram based on the GMF theory
in the $2\times2$ unit cell (see Fig. \ref{fig:triorbcur}).
Large scale GMF calculations in a $30\times 30$ lattice are
performed to confirm the stripe ordered superfluid (SF) phase
at points 1, 2 and 3 with $(t/U,\mu/U)=(0.02,2),
(0.03,1.5)$ and $(0.038,2.2)$, respectively.
B) The flux $\Phi$ around a rhombic plaquette v.s. $t/(nU)$.
It decays from $\frac{1}{3}$ in the strong coupling limit to $0$ 
in the non-interaction limit.
The solid line is the GMF result at $n=3$, while the dashed line
is based on the energy function Eq. \ref{eq:tricond} of the
trial condensate.
}
\label{fig:phase}
\end{figure}

The stripe order persists in the strong
coupling SF regime due to the interaction among vortices.
Consider a plaquette with vorticity $+\frac{1}{6}$, thus
its three vertices are  with two $+1$'s and one $-1$.
The neighbouring plaquette sharing the edge with two $+1$'s must have the
same vorticity, and merges with the former one to form a rhombic plaquette
with the total vorticity $+\frac{1}{3}$.
Thus the ground state should exhibit a staggered pattern of rhombic
plaquettes with vorticity of $\pm \frac{1}{3}$.
The stripe pattern of OAM moments is the only possibility to
satisfy this requirement.  The stripe phase obtained here is quite
general: for example, it will 
also appear in the triangular-lattice $p+ip$ Josephson-junction array
~\cite{moore2004} if tunneling is dominated by the
momentum-reversing process~\cite{castelnovo2004}, rather than
by the momentum-conserving process that gives a uniform state.
This stripe ordering possibility in the triangular $p$+$ip$
Josephson junction arrays has not been earlier appreciated.

In addition, the stripe order  results in the staggered bond currents.
We further optimize the $U(1)$ phase variables, and find that
their pattern is the same as that in the weak coupling limit.
Taking into account the equal weight of $p_{x,y}$ in each site,
we find  no phase mismatch along each horizontal bond,
but a phase mismatch of $\Delta\theta=\frac{\pi}{6}$ along each tilted bond.
As a result, on each bond around the  rhombic plaquette, the Josephson
current is $j=\frac{t n_0}{2} \sin \Delta \theta$
where $n_0/n$ is the condensate fraction,
and the current direction is specified by arrows in Fig. \ref{fig:triorbcur}.
The total phase winding around each rhombic plaquette
is $4\Delta\theta=\frac{2}{3} \pi$, and thus agrees with the vorticity
of $\frac{1}{3}$.
%This staggered plaquette bond current  is similar to the
%$d$-density wave state proposed in high-T$_c$ cuprates \cite{chakravarty2001},
%but with a completely different microscopic origin.
We emphasize that the $p$-band square lattice case does not have this 
interesting physics \cite{liu2006}.

Since the stripe order exists in both strong and weak coupling limits, 
it should also exist at intermediate coupling strength.
We have confirmed this conjecture using the Gutzwiller mean field (GMF) 
theory in a $30\times 30$ lattice for three systems (marked as points
1, 2, and 3 in Fig. \ref{fig:phase}).
We find that the stripe ordered ground state 
is stable against small perturbations   in all three cases.
We further apply the GMF theory to the $2\times 2$ unit cell (Fig.
\ref{fig:triorbcur}), and obtain the phase diagram of the stripe ordered
SF  and MI phases (Fig.~\ref{fig:phase}A).
To understand the GMF numerical results, we write the trial
condensate with the $p$-orbital configuration on each site as
$ e^{i\phi_{\vec r}} (\cos\alpha |p_x\rangle + i \sigma_{\vec r}
\sin\alpha |p_y\rangle)$.
It turns out that the pattern for the $U(1)$ phase does not
depend on $\alpha$, and remains the same for all the coupling
strength.
The phase mismatch $\Delta \theta$ on the tilted bonds reads
$\Delta\theta=2\gamma-\pi/2$ with $\tan \gamma={\sqrt 3} \tan \alpha$,
and  the corresponding Josephson current is
$j= n_0 t \sin \Delta \theta$.
The value of $\alpha$ is determined by the minimization of the
energy per particle of the trial condensate as
\bea
{\cal E}(\alpha) = -t [1+ 2 \sin (2 \alpha +\frac{\pi}{6})]- 
\frac{n U}{6} \sin^2 2 \alpha+\frac{n U}{3}.
\label{eq:tricond}
\eea
In the strong (weak) coupling limit, the energy minimum 
is located at $\alpha=\frac{\pi}{4} ~(\frac{\pi}{6}) $, and thus
the flux in each rhombic plaquette 
$\Phi= 4\Delta \theta/(2\pi)=0~ (\frac{1}{3})$, which agrees with the previous
analyses.
For the intermediate interaction, we present both results of $\Phi$
at $n=3$ based on  the GMF theory and Eq.(\ref{eq:tricond}) in
Fig.~\ref{fig:phase}B.
They agree with each other very well, and confirm the validity of
the trial condensate.
Moreover, in the momentum space, the trial condensate for a general 
$\alpha$  can be expressed as
$\frac{1}{\sqrt {N_0 !}}\Big\{\frac{1}{\sqrt 2} (\psi^{\prime\dagger}_{K_2} 
+i \psi^{\prime\dagger}_{K_3})\Big\}^{N_0}|0\rangle,
\label{eq:generalcond}$
where $\psi^\prime_{K_{2,3}}(\vec r)= 
e^{i\vec K_{2,3} \cdot \vec r} |\phi_{2,3}(\alpha)\rangle$
with $|\phi_{2,3}(\alpha)\rangle= -\cos \alpha |p_x\rangle\mp \sin \alpha
|p_y\rangle$ respectively.

The OAM is of the Ising type, thus the stripe ordering
is robust against small perturbations such as
a small value of the $\pi$-type bonding $t_\perp\ll t_\pp$.
Further, we also check that the phase pattern in  Fig. 
\ref{fig:triorbcur} remains unchanged from minimizing 
the ground state energy.
In particular, in the weak-coupling limit, the band minima and 
corresponding eigenvectors do not change at all in the
presence of a small $t_\perp$, which renders the
above conclusion obvious.

The formation of the on-site orbital moment does not depend
on the inter-site phase coherence, and thus the Ising variables
can be ordered even in the MI state.
We performed a ring exchange analysis
showing the existence of the stripe-ordering  at $n\ge 2$ 
provided the screening length of the interaction among vortices
is larger compared to the size of the
four-site plaquette.
Due to loss of the inter-site phase coherence,
bond currents disappear in the MI phase.

\begin{figure}
\vspace{-0.5cm}
\centering\epsfig{file= 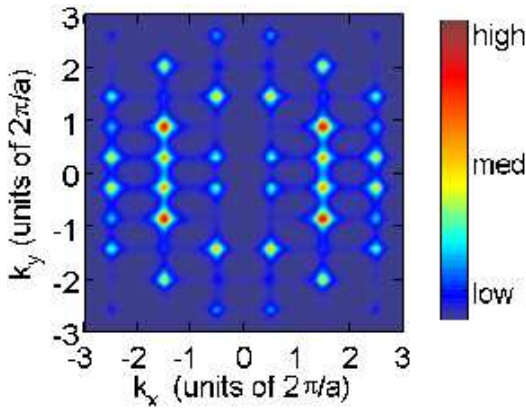,clip=1,width=0.8\linewidth,angle=0}
\caption{Predicted TOF image for the stripe-ordered superfluid phase
with the condensation wavevectors $K_{2,3}$. Note that the locations
of the highest peaks depend on the size $l_{x,y}$ of the $p$-orbital
Wannier function. Parameters are: $\alpha={\pi/6}$; $l_{x,y}/a=0.1$;
the $\delta$-function is replaced by a Lorentzian line for
display. 
%{\bf Note: have added $\alpha=...$?!} 
}
\label{fig:tof}
\end{figure}

Our predicted stripe phase should manifest itself in the time of flight 
(TOF) signal as depicted in Fig.  \ref{fig:tof}.
In the SF state, we assume the stripe ordering wavevector $K_1$,
and the corresponding condensation wavevectors at $K_{2,3}$.
As a result, the TOF density peak position after a fight time of $t$
is shifted from the reciprocal lattice vectors $\vec G$ as follows
\bea
\avg{n(\vec r)}_t &\propto& \sum_{\vec G}\Big\{ |\phi_2 (\alpha, \vec k)|^2 
\delta^2(\vec k-\vec K_2- \vec G)\nn\\
&+&|\phi_3 (\alpha, \vec k)|^2 \delta^2(\vec k-\vec K_3-\vec G)\Big\},
\eea
where $\vec k=m \vec r /( \hbar t)$; $\phi_{2,3}(\alpha,\vec k)$ is
the Fourier transform of the Wannier $p$-orbital wavefunction 
$|\phi_{2,3}(\alpha)\rangle$,
and  $\vec G = \frac{2\pi}{a}[m, (-m+2n)/\sqrt 3]$ with $m,n$ integers.
Thus Bragg peaks should occur at $\frac{2\pi}{a}
[m\pm\frac{1}{2}, \frac{1}{\sqrt 3} (-m+2n+\frac{1}{2})]$.
Due to the form factors of the $p$-wave Wannier orbit wavefunction
$|\phi_{2,3}(\alpha, \vec k)|^2$, the locations of the highest peaks 
is not located at the origin but around $|k|\approx 1/l_{x,y}$.
Due to the breaking of lattice rotation symmetry,
the pattern of Bragg peaks can be rotated at angles of
$\pm \frac{2\pi}{3}$.
In the MI phase, Bragg peaks disappear due to the loss of phase coherence.
Instead, the stripe order 
appears in the noise correlations $\avg{n(\vec r) n (\vec r^\prime)}$,
which  exhibit not only the usual peaks at $\vec G$, but also peaks
located at $\vec K_1+\vec G$.

In summary, optical lattices provide a promising direction 
to study new phases of orbital physics.
We focus on the $p$-band bosons in the frustrated 
triangular lattice which exhibits novel stripe orbital ordering
of the onsite OAM due to the geometric frustration effect
to the phase coherence.
In the SF phase, the staggered plaquette bond currents
are also induced, reminiscent
of the $d$-density wave  proposal for the pseudogap phase
in high T$_c$ cuprates \cite{chakravarty2001} but
with completely different microscopic origin.
The stripe order persists in the MI phase 
even with the loss of superfluidity. 
The pattern of the stripe order can be observed in the TOF experiment.
The orbital physics of cold atoms
opens up intriguing problems yet to be explored, 
for example, the $p$-orbital ordering in other frustrated lattices
such as the Kagome.

C. W. thanks L. M. Duan and T. L. Ho for helpful discussions.  C.W. is
supported by the the NSF Phy99-07949. W. V. L. is supported in part by
ORAU Ralph E. Powe Award.  J. E. M. is supported by NSF DMR-0238760.
S. D. S. is supported by LPS-NSA.


\begin{thebibliography}{10}  


\bibitem{myatt1997} 
D. M. Stamper-Kurn {\it et al}, \PRL {\bf 80}, 2027 (1998);
M. S. Chang {\it et al.}, Nature Physics,
{\bf 1}, 111 (2005).

\bibitem{ho1998} T. L. Ho, {\it et al.}, \PRL {\bf 81}, 742(1998);
T. Ohmi {\it et al.}, J. Phys. Soc. Jpn., 67,1822 (1998);
E. Demler {\it et al}, \PRL {\bf 88}, 163001(2002); 
F. Zhou, \PRL {\bf 87}, 80401(2001).


\bibitem{wu2003} C. Wu {\it et al.}, \PRL {\bf 91}, 186402 (2003).

\bibitem{wu2005} C. Wu {\it et al.}, cond-mat/0512602;
T. L. Ho {\it et al.}, \PRL {\bf 82}, 247 (1999).

\bibitem{wu2005a} C. Wu, \PRL {\bf 95}, 266404 (2005);
P. Lecheminant {\it et al.}, Phys. Rev. Lett. 95,  240402 (2005).


\bibitem{tokura2000} Y. Tokura and N. Nagaosa, Science {\bf 288},
462 (2000).

\bibitem{browaeys2005} A. Browaeys {\it et. al}, \PRA {\bf 72}
53605 (2005).

\bibitem{kohl2005} M. Kohl {\it et. al}, \PRL {\bf 94}, 80403 (2006).

\bibitem{scarola2005} V. W. Scarola {\it et. al},
\PRL~ {\bf 65}, 33003 (2005).

\bibitem{isacsson2005} A. Isacsson {\it et. al}, \PRA~ {\bf 72},
053604 (2005).

\bibitem{liu2006} W. V. Liu and C. Wu, cond-mat/0601432.

\bibitem{kuklov2006} A. B. Kuklov, cond-mat/0601416.

\bibitem{moore2004} J. E. Moore {\it et al.}, \PRB~ {\bf 69},
104511 (2004).

\bibitem{xu2004} C. Xu {\it et al.}, \PRL~{\bf 93},
47003(2004).

\bibitem{nussinov2005}  Z. Nussinov, {\it et. al},
Phys. Rev. B {\bf 71}, 195120 (2005).

\bibitem{jaksch2005}  D. Jaksch and P. Zoller,
Ann. Phys. {\bf 315}, 52 (2005).

\bibitem{mori1998} S. Mori {\it et al.}, Nature {\bf 392}, 473 (1998).

\bibitem{kivelson1998}J. Zaanen {\it et al.}, \PRB ~{\bf 40},
7391 (1989);  S. A. Kivelson, {\it et al.},
Rev. Mod. Phys. {\bf 75}, 1201 (2003).

\bibitem{lilly1999} M. P. Lilly {\it et al.}, \PRL~ {\bf 82}, 394 (1999)
;R. R. Du {\it et al.}, Solid State Commun.
{\bf 109}, 389 (1999).

\bibitem{castelnovo2004} C. Castelnovo {\it et al.}, \PRB~ {\bf 69},
104529 (2004).




\bibitem{chakravarty2001} S. Chakravarty {\it et al.}, \PRB~{63},
94503 (2001).

\end{thebibliography}
\end{document}